\documentclass[prl,twocolumn,superscriptaddress,showpacs]{revtex4}
\usepackage{bm}
\usepackage{amssymb}
\usepackage{graphicx}
\newcommand{\omspin}{\Omega^{\mbox{\footnotesize spin}}_1}
\begin{document}
\title{Measurement of gravitational spin-orbit coupling in a binary pulsar system}
\date{\today}
\author{I. H. Stairs}
\affiliation{Department of Physics and Astronomy, University of British Columbia, 6224 Agricultural Road,
Vancouver, BC V6T 1Z1, Canada}
\email{stairs@astro.ubc.ca}
\homepage{http://www.astro.ubc.ca/people/stairs/}
\author{S. E. Thorsett}
\affiliation{Dept.\ of Astronomy \& Astrophysics, University of California, 
Santa Cruz, CA 95064}
\author{Z. Arzoumanian}
\affiliation{USRA, Laboratory for High-Energy Astrophysics, NASA-GSFC, Code 662, Greenbelt, MD 20771}
\begin{abstract}
In relativistic gravity, a spinning pulsar will precess as it orbits a
compact companion star. We have measured the effect of such precession
on the average shape and polarization of the radiation from
PSR~B1534+12. We have also detected, with limited precision,
special-relativistic aberration of the revolving pulsar beam due to
orbital motion. Our observations fix the system geometry, including
the misalignment between the spin and orbital angular momenta, and
yield a measurement of the precession timescale consistent with the
predictions of General Relativity.
\end{abstract}
\maketitle

\section{1. Introduction}

Parallel transport of the angular momentum of a gyroscope moving
in curved spacetime leads to geodetic precession \cite{des16}. In the
solar system, the only observed example is precession of the
Earth-Moon system as it orbits the Sun \cite{wnd96}. The recently
launched Gravity Probe B experiment plans to measure the geodetic
precession of a gyroscope in Earth orbit as well as its
gravitomagnetic Lense-Thirring precession \cite{bep00}.

Binary pulsar systems are important laboratories for gravitational physics
\cite{dt92}, in part because the strong self-gravity of neutron stars 
($GM/Rc^2\sim0.2$) raises the possibility of deviations as large as
order unity in some alternate gravity theories---even theories that
agree with general relativity (GR) in weak-field tests
\cite{de96}.  Immediately after the discovery of the first binary
pulsar, PSR~B1913+16 \cite{ht75a}, it was realized that geodetic
precession could lead to variations in the path of the observer's line
of sight across the pulsar's magnetic pole, and hence changes in the
radiation pattern at Earth \cite{dr74}. Recently, the anticipated
pulse-profile variations have been observed, in good qualitative
agreement with predictions \cite{wrt89,kra98,wt02}, but uncertainties
in the intrinsic beam shape prevent a quantitative measurement of the
precession rate.

Here we describe observations of the double-neutron-star system
PSR B1534+12, in which precession is changing the observed pulsar
profile by about 1\% per year \cite{arz95,stta00}. We outline a new,
general technique for combining measurements of long-term quasisecular
variations caused by geodetic precession with those of periodic
variations induced by special relativistic aberration modulated by
orbital motion. Together, the observations allow a quantitative
estimate of the precession rate, independent of the unknown pulsar
beam shape. Although the precision is still low, we show that this
model-independent precession rate is consistent with the predicted
rate in GR.  The polarization properties of the pulsar signal are also
changing: with the additional assumption that the magnetic field
structure is dipolar, we can determine the angles between the spin and
orbital angular momenta and the line of sight to the pulsar.

\begin{figure}
\includegraphics[width=2.5in]{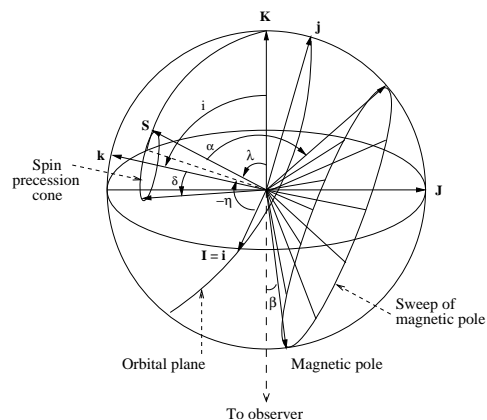}
\caption{\label{fig:geom}  The spin and orbital geometry we derive
for PSR~B1534+12, after \cite{dt92}.  The vectors $\mathbf{I}$ and
$\mathbf{J}$ denote the plane of the sky, while $\mathbf{i}$ and
$\mathbf{j}$ show the plane of the pulsar orbit.  The pulsar spin axis
at the current epoch is $\mathbf{S}$ and the orbital angular momentum
direction is $\mathbf{k}$; the precession cone with opening
(misalignment) angle $\delta$ is shown. The angle between $\mathbf{S}$
and the line of sight is $\zeta$; the supplemental angle
$\lambda\equiv\pi-\zeta$ is shown.  The angle between $\mathbf{S}$ and
the magnetic pole is $\alpha$, and $\beta=\zeta-\alpha$ is the minimum
impact angle of the magnetic pole on the line of sight. A second cone
indicates the sweep of the magnetic pole at the current epoch.  The
projection of $\mathbf{S}$ on the plane of the sky (indicated by the
dashed line) provides angle $\eta$, measured counterclockwise from the
ascending node; $-\eta$ is shown.  Spherical geometry gives $\cos
\delta = - \sin i \sin \lambda \sin \eta + \cos \lambda \cos i$.  }
\end{figure}

The spin and orbital geometry is shown in Fig.~\ref{fig:geom}.  The
pulsar spin axis $\mathbf{S}$ will precess around the total angular
momentum, which is very well approximated by the orbital angular
momentum direction $\mathbf{k}$.  This precession will cause
potentially detectable periodic variations of the projection of
$\mathbf{S}$ on the plane of the sky, and on the inclination of
$\mathbf{S}$ with respect to the observer.  In GR, the time averaged
precession rate of the pulsar can be written
\cite{dr74}:
\begin{equation} 
\label{eqn:omspin}
\omspin= {\frac{1}{2}} {\left( \frac{GM_{\odot}}{c^3} \right)} ^{2/3} {\left( \frac{P_{\rm b}}{2 \pi} \right)} ^{-5/3}
\frac{m_2(4m_1+3m_2)}{(1-e^2)(m_1+m_2)^{4/3}},
\end{equation}
where $G$ is Newton's constant, $M_{\odot}$ is the mass of the Sun,
$c$ is the speed of light, $m_1$ and $m_2$ are the pulsar and
companion masses, respectively, $P_{\rm b}$ is the orbital period, and
$e$ is the eccentricity.  A description in generalized theories of
gravity is given in \cite{dt92}.  For PSR~B1534+12, using the stellar
masses determined through high-precision timing \cite{sttw02}, the
precession rate predicted by GR is $0.51^{\circ}\,$/yr.

\section{2. Observations and Analysis}

Observations were made with the 300-m Arecibo radio telescope, using
the ``Mark~IV'' data acquisition system \cite{sst+00} at an observing
frequency of 430\,MHz.  The signal was processed using coherent
dedispersion, providing full polarization information as well as a
pulse shape unaffected by dispersive smearing in the interstellar
medium.  Data acquisition details have been described elsewhere
\cite{sttw02}.

The data span the interval from mid-1998 to mid-2003, incorporating
some 400 hours of observing time.  The pulsar was observed biweekly or
monthly. The observed signal strength varied widely because of
interstellar scintillation; only epochs with high signal-to-noise
ratio were used here.  Campaigns of roughly 12 contiguous observing
days were also conducted every summer except 2002.

\begin{figure}
\includegraphics[width=3in]{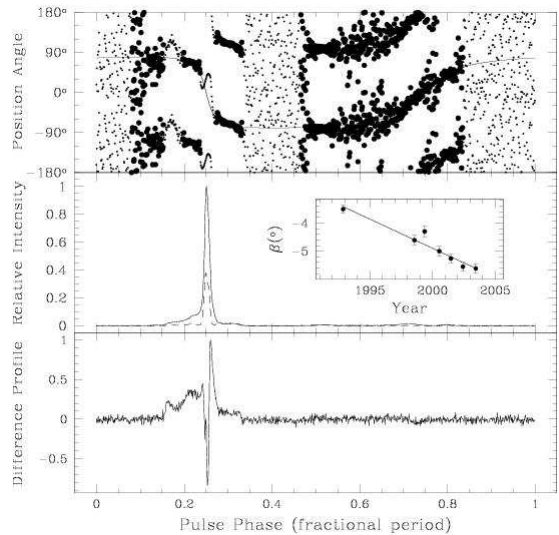}
\caption{\label{fig:pol} Top panel: the position angle of
linear polarization in 2001 June, measured clockwise on the plane of
the sky (the convention in \cite{dt92}), with best fit rotating vector
model (RVM) overlaid.  Only the position angle points indicated by
large dots were used in the RVM fit; these were weighted by their
uncertainties, with a small uncertainty added in quadrature to account
for deviations from the RVM.  Middle panel: total intensity (solid)
and linear polarization (dashed) profiles in 2001 June.  This profile
is very similar in shape to our ``reference'' profile $P_0$. Inset:
evolution of impact angle $\beta$ with time.  Bottom panel:
``Difference'' profile $P_1$, representing essentially
the time-derivative of the observed profile.}
\end{figure}

The cumulative pulse profile from the representative 2001 June epoch
is shown in Fig.~\ref{fig:pol}.  Here we also show a fit of the linear
polarization to the standard ``rotating vector model'' (RVM)
\cite{rc69a}, in which the position angle of linear polarization
$\psi$ is assumed to be parallel to the plane of curvature of magnetic
dipole field lines rotating with the star, giving
\begin{equation}
\label{eqn:rvm}
\tan\left[\psi\left(\phi\right)-\psi_{0}\right]=\frac{\sin\alpha\sin\left(\phi-\phi_0\right)}{\cos\alpha\sin\zeta-
\sin\alpha\cos\zeta\cos\left(\phi-\phi_0\right)},
\end{equation}
where $\phi$ is the pulse phase, $\phi_0$ and $\psi_0$ are constants,
$\alpha$ is the magnetic inclination angle, and
$\zeta$ is the angle between $\mathbf{S}$ and the line of sight.

The position angle sweep is observed over most of the pulsar period,
and the model fit is generally good. There are strong deviations from
the model near the pulse peak, as is often seen for ``core'' profile
components \cite{ran83}. We exclude this region from our fits. The
data are consistent with a roughly orthogonal rotator model, with
$\alpha=102.8\pm0.5^{\circ}$ and the line of sight passing between the
magnetic pole and the stellar equator, within a few degrees of the
magnetic pole.  We also consider the time evolution of $\alpha$ and
the impact parameter of the line of sight on the magnetic pole
$\beta$, using cumulative profiles from each campaign, one especially
strong biweekly observation, and an earlier coherently-dedispersed
profile from observations with the ``Mark~III'' data acquisition
system \cite{aptw96}.  As expected, the data are consistent with no
evolution of $\alpha$.  However, as shown in Fig.~\ref{fig:pol},
$\beta$ is changing with time, at a rate
$d\beta/dt=-0.21\pm0.03^\circ$/yr, resulting in a larger impact
parameter at later times.  This change in $\beta$ is direct evidence
of geodetic precession, and can be related to the system geometry and
precession rate in a simple fashion \cite{dt92}:
$d\beta/dt=\omspin\cos\eta\sin i$ (see Fig.~\ref{fig:geom}).

The {\it shape} of the profile is also changing with time and orbital
phase, allowing a completely independent probe of the precession.
Secular changes in the profile were first noticed at 1400\,MHz
\cite{arz95}, but evolution of the 430-MHz emission only
became apparent with coherently dedispersed observations
\cite{stta00}.  Shape variations are more difficult to connect directly 
to the precession rate than polarization changes. It can be done with
an assumed model of the beam shape, as has been attempted for
PSR~B1913+16 \cite{kra98,kra02,wt02}, but in that case the true beam
shape is still debated and the results are therefore less than
satisfactory.

Here we note that it is possible to make a model-independent
precession estimate by also measuring the orbital modulation of the
profile shape caused by aberration---a special-relativistic effect
independent of strong-field gravity.  Aberration shifts the observed
angle between the line of sight and spin axis by an amount
\cite{dt92}:
\begin{equation}
\label{eqn:deltaazeta}
\delta_{\rm A}\zeta=\frac{\beta_1}{\sin i}\left[-\cos\eta\,
S\left(u\right)+\cos i\sin\eta\,C\left(u\right)\right],
\end{equation}
\noindent where $\beta_1\equiv nx/\sqrt(1-e^2)$ is the characteristic 
velocity of the pulsar, with the orbital frequency $n\equiv 2\pi/P_b$,
the projected semimajor axis $x\equiv a_1\sin i/c$, and the
eccentricity $e$ all available from timing data, and where
$C(u)\equiv\cos[\omega+A_e(u)]+e\cos\omega$ and
$S(u)\equiv\sin[\omega+A_e(u)]+e\sin\omega$ are functions of the
time-dependent angle of periastron passage $\omega$ and the eccentric
anomaly $u$ through the true anomaly
$A_e\left(u\right)\equiv
2\arctan\left[\left(\frac{1+e}{1-e}\right)^{1/2}\tan\frac{u}{2}\right]$.

Now let ${F}(\zeta)$ be any function defined on the observed
pulsar signal that depends on the viewing angle (such as integrated
intensity, component width, polarization fraction, etc.). For small
changes in the impact parameter, we Taylor-expand
${F}(\zeta)\approx{F}(\zeta_0)+\zeta{F}'$
where prime denotes derivative with respect to $\zeta$. The effects of
aberration and precession can then be written
\begin{eqnarray}
\delta_{\rm A}{F}&=&{F}'\frac{\beta_1}{\sin i}
\left[-\cos\eta\,S\left(u\right)
+\cos i\sin\eta\,C\left(u\right)\right],\\ 
\frac{d{F}}{dt}&=&{F}'\omspin\sin i\cos\eta. \label{eqn:aberr}
\end{eqnarray}
\noindent The unknown beam shape enters only through ${F}'$, 
which can be eliminated by dividing these two equations. Measurements
of both the orbital variation of ${F}$ and its secular drift
thus allow $\tan \eta$ and $\omspin$ to be determined in a model
independent way.

\begin{figure}
\includegraphics[width=3in]{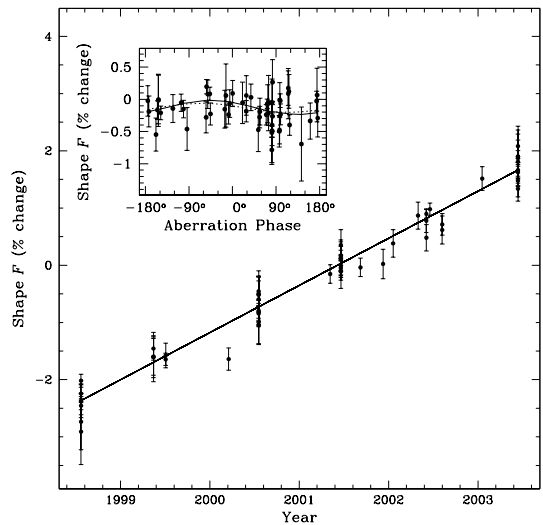}
\caption{\label{fig:fit}  
The shape parameter ${F}=c_1/c_0$ (see text) is shown as a function of
date in the main panel and aberration phase (essentially the true
anomaly corrected for the advance of periastron) in the inset.  The
best-fit model is shown by the solid line in each panel, and in the
orbital-phase plot, the GR prediction based on the RVM model is
indicated by the dotted line.  We have included a small error in
quadrature with the measurement errors to account for smearing caused
by averaging over a range of orbital phase.  To account for systematic
errors, variable data quality, and uneven time sampling, we have used
a bootstrap analysis \cite{ptvf92} to estimate the uncertainties on
model parameters. The resulting values and uncertainties are in good
agreement with estimates obtained by scaling measurement errors to
obtain a reduced-$\chi^2$ of 1.}
\end{figure}

For PSR~B1534+12, we have measured the evolution of the total
intensity profile (Fig.~\ref{fig:fit}).  The strongest data scans from
the annual campaigns were averaged into 12 orbital phase bins, and
analyzed together with the strong biweekly scans.  Profiles with
unusually low signal-to-noise ratios or suspect calibrations were
discarded.  We used standard principal component (PC) analysis
techniques \cite{ptvf92} to derive orthogonal ``reference'' ($P_0$)
and ``difference'' ($P_1$) profiles that completely described the
profile evolution; $P_1$ and a single-epoch profile very similar to
$P_0$ are shown in Fig.~\ref{fig:pol}.  The pulse profile $P$ of each
observation is well modeled as a linear combination $P=c_0P_0+c_1P_1$.
Because the overall amplitude varied with scintillation, we chose as
our observable quantity ${F}$ the ratio $c_1/c_0$.  The PC analysis
provided estimates of $c_0$, $c_1$, and their uncertainties, which we
independently checked through a frequency-domain cross-correlation
technique in which a linear combination of the two profiles was fit in
an iterative manner.  Finally, we simulated the cross-correlation
analysis to assess its sensitivity to systematic errors induced by
imperfect calibration or polarization cross-coupling, finding that
such problems should be negligible in our dataset.

The secular trend in ${F}$ is evident in Fig.~\ref{fig:fit}.  This
corresponds to a decrease in the intensity of the core region of the
profile relative to the lower level emission in the wings, consistent
with early indications from 1400\,MHz data, and as expected if
precession is moving our line of sight away from the magnetic pole, as
indicated by the polarization analysis above.  The residuals after
removing the best fit line are shown as a function of orbital phase. A
simultaneous linear fit of ${F}$ as a function of date and of $S(u)$
and $C(u)$ gives the constraints
$\omspin\sin^2i=0.42^{+0.46}_{-0.15}\,^{\circ}\,$/yr and $\omspin\sin
i\tan i\cot\eta=0.42^{+0.49}_{-0.16}\,^{\circ}\,$/yr (68\%
confidence), where we have used $\beta_1=0.67\times10^{-3}$.  A fit of
${F}$ only as a function of date yields a $\chi^2$ value that is 15\%
higher than that for the full fit.

\section{3. Discussion}

We have observed both long- and short-term variations in the pulse
shape of PSR B1534+12, as expected from geodetic precession and
aberration.  Assuming a dipolar field geometry and GR, the impact
parameter change $d\beta/dt=\omspin\cos\eta\sin
i=-0.21\pm0.03^{\circ}/$yr yields a measurement of the previously
unknown angle $\eta=\pm115.0\pm3.8^\circ$.  Using the pulse timing
value $\sin i=0.975$ \cite{sttw02} but making no assumptions about the
validity of the RVM, our measured pulse profile variations yield,
using equations~4~and~5, the consistent result $\eta =
\pm103\pm10^\circ$.  Moreover, we may now solve for the precession
rate, $\omspin=0.44^{+0.48}_{-0.16}\,^\circ$/yr (68\% confidence) or
$\omspin=0.44^{+4.6}_{-0.24}\,^\circ$/yr (95\% confidence). This value
compares well with the GR-predicted rate of $\omspin=0.51^\circ$/yr.

The misalignment angle $\delta$ between the spin and orbital angular
momenta can also be constrained. The angle $\lambda$ is known from the
polarization studies. Only the absolute values of $\sin \eta$ and
$\cos i$ are known, but our profile fit requires that $\cos i \tan
\eta > 0$.  Therefore there are two possible geometries: $i=77.2^\circ$ 
and $\eta=-115^\circ$, which gives $\delta=25.0\pm3.8^\circ$, or
$i=102.8^\circ$ and $\eta =115^\circ$, which gives
$\delta=155.0\pm3.8^\circ$.  Both give identical results for
precession in GR. As the angular momenta were almost certainly aligned
before the second supernova, the smaller misalignment value is favored
on astrophysical grounds \cite{bai88}.  The preferred geometry then
has $i=77.2^\circ$, $\eta=-115.0\pm3.8^\circ$, and
$\delta=25.0\pm3.8^\circ$ (Fig~\ref{fig:geom}).  The misalignment
angle can be used to constrain mass loss and asymmetry in the second
supernova.  A full analysis will be published elsewhere, along with a
study of the two-dimensional beam geometry of PSR~B1534+12.

Although the precision is, as yet, limited, this is the first
beam-model-independent measurement of the precession rate of a binary
pulsar and reconstruction of the full three dimensional geometry of a
binary pulsar system, including the misalignment angle.  Future
prospects for improvement include direct estimation of the angle
$\eta$ by combining scintillation studies \cite{bplw02} with
polarimetry \cite{dt92}.  The derived geometry also allows us to
predict the effects of aberration on the pulse timing \cite{dt92};
this will in principle allow more precise timing tests of GR in
future.

We emphasize that the general technique of combining observations on
the orbital and precessional timescales to make model independent
precession rate estimates is potentially far more general than the
particular example given here.  An especially interesting prospect is
the recently discovered highly relativistic system PSR~J0737$-$3039
\cite{bdp+03}; with $\beta_1$ nearly twice as large as that of
B1534+12, and a predicted precession timescale for the recycled pulsar
of only 75 years, both effects will be quickly measured for this new
system.

\begin{acknowledgments}
The Arecibo Observatory is operated by Cornell University under a
cooperative agreement with the NSF.  IHS is supported by an NSERC
Discovery Grant.  SET acknowledges support from the NSF.  ZA is
supported by a NASA grant.  We thank D. Nice for his contributions to
the software that made this work possible, J. Taylor for significant
help and encouragement in the early stages, and numerous colleagues
for assistance with observations.  We also thank the Arecibo
Scheduling Advisory Committee for its longstanding support of this
project.
\end{acknowledgments}


\end{document}